\documentclass[prd,superscriptaddress,nofootinbib,showpacs,preprint]{revtex4}
\usepackage{amsmath}
\usepackage{graphicx}
\usepackage{hyperref}  
\usepackage{color} 
\usepackage{epsfig}
\usepackage{epsf}

\begin{document}
\title{Type III Seesaw and Dark Matter in a Supersymmetric Left-Right Model \\}
\author{Debasish Borah}
\email{debasish@phy.iitb.ac.in}
\affiliation{Indian Institute of Technology, Bombay, Mumbai - 400076, India}
\affiliation{Indian Institute of Technology Gandhinagar, Ahmedabad - 382424, India}
\author{Sudhanwa Patra}
\email{sudhakar@prl.res.in}
\affiliation{Physical Research Laboratory, Ahmedabad-380009, India}

\begin{abstract}
We propose a new supersymmetric left right model with Higgs doublets carrying odd B-L charge, 
Higgs bidoublet and heavy Higgs triplets with zero B-L charge and a set of sterile neutrinos 
which are singlet under the gauge group. We show that spontaneous parity violation can be 
achieved naturally in this model and the neutrino masses arise from the so called type III 
seesaw mechanism. We also discuss the possible phenomenology in the context of neutrino masses
and dark matter.
\end{abstract}

\maketitle

\section{Introduction}
Despite the success of the standard model(SM) there are still unsolved 
problems in particle physics which motivate searches for more fundamental 
theories. One of the most appealing extensions of the SM is the left-right 
symmetric model based on gauge group $SU(2)_L \times SU(2)_R \times U(1)_{B-L}$ where parity is spontaneously broken and smallness of neutrino masses arises naturally. However like the SM, this left-right symmetric model also suffers from the hierarchy problem: the masses of the Higgs scalar diverge quadratically.
As in SM, supersymmetric(SUSY) counterpart of the left-right model cures this hierarchy problem. Hence left-right supersymmetric model \cite{Pati:1974yy,
Mohapatra:1974gc, Senjanovic:1975rk, Mohapatra:1980qe, Deshpande:1990ip}
becomes an appealing model which can cure many problems in the standrad model.

 Among many attractive features of this model is the capability to explain the 
smallness of the neutrino mass via so called seesaw mechanism. Due to inclusion 
of right handed neutrino states as a principle, such models 
provide a natural explanation for the smallness of
neutrino masses \cite{Fukuda:2001nk, Ahmad:2002jz, Ahmad:2002ka, Bahcall:2004mz} 
via see-saw mechanism \cite{Minkowski:1977sc, GellMann:1980vs, Yanagida:1979as, Mohapatra:1979ia}.
This class of models also provide a natural embedding of electroweak
hypercharge, giving a physical explanation for the required extra $U(1)$ as
being generated by the difference between the 
baryon number ($B$) and the lepton number ($L$). 

The minimal SUSY left-right model have been reviewed extensively where the number 
of Higgs fields is the smallest possible. It is found that minimal model suffers from problem with spontaneous parity breaking as we expect parity breaking at low 
energy theory.  Recently, the spontaneous parity violation was shown in an extension of 
supersymmetric Left-Right model \cite{Patra:2009wc} where the extra field added to the minimial field content is a
bitriplet under $SU(2)_L \times SU(2)_R$.
\color{black} There are many models which can achieve spontaneous parity breaking by adding 
more fileds to minimal SUSYLR field content. In the next section we discuss this minimal SUSYLR model and its 
extension by adding two extra Higgs triplets ~\cite{Aulakh:1998nn, Aulakh:1997ba}. 
Then we describe our model with Higgs doublets instead of triplets and discuss the 
possible phenomenology associated with it.

\section{Minimal SUSYLR model}
\label{Aulakh}
The minimal set of the Higgs fields in the non-SUSY model consists of of a 
bidoublet $\Phi_u$ and a $SU(2)$ triplet $\Delta$. In the supersymmetric version, 
the cancellation of chiral anomalies among the fermionic partner of the triplet 
Higgs fields $\Delta$ requires introduction of the second triplet $\bar{\Delta}$ 
with opposite $U(1)_{B-L}$ quantum number. Due to the conservation of the B-L symmetry, 
$\Delta$ does not couple to the leptons and quarks. Another bidoublet $\Phi_d$ 
is introduced to avoid the trivial Kobayashi-Maskawa matrix for quarks. This 
because supersymmetry forbids a Yukawa coupling where the bidoublet appears 
as a conjugate. 
The complete structure of the field content of the minimal supersymmetric left right model is 
given by
\begin{equation}
Q=
\left(\begin{array}{c}
\ u \\
\ d
\end{array}\right)
\sim (3,2,1,\frac{1}{3}),\hspace*{0.8cm}
Q_c=
\left(\begin{array}{c}
\ d_c \\
\ u_c
\end{array}\right)
\sim (3^*,1,2,-\frac{1}{3}),\nonumber 
\end{equation}
\begin{equation}
L=
\left(\begin{array}{c}
\ \nu \\
\ e
\end{array}\right)
\sim (1,2,1,-1), \quad
L_c=
\left(\begin{array}{c}
\ \nu_c \\
\ e_c
\end{array}\right)
\sim (1,1,2,1)
\end{equation}
where the numbers in the brackets denote the quantum numbers under 
$SU(3)_c \otimes SU(2)_L \otimes SU(2)_R \otimes U(1)_{B-L}$. 
Also here the convention is such that $L\rightarrow U_LL$ under 
$SU(2)_L$, but $L^c\rightarrow U^*_RL^c$ under $SU(2)_R$. 
Unlike in MSSM here the Higgs sector consists of the bidoublet 
and triplet superfields:
\begin{equation}
\Phi_1=
\left(\begin{array}{cc}
\ \phi^0_{11} & \phi^+_{11} \\
\ \phi^-_{12} & \phi^0_{12}
\end{array}\right)
\sim (1,2,2,0),\hspace*{0.2cm} 
\Phi_2=
\left(\begin{array}{cc}
\ \phi^0_{21} & \phi^+_{21} \\
\ \phi^-_{22} & \phi^0_{22}
\end{array}\right)
\sim (1,2,2,0), \nonumber 
\end{equation}
\begin{equation}
\bigtriangleup =
\left(\begin{array}{cc}
\ \delta^+_L/\surd 2 & \delta^{++}_L \\
\ \delta^0_L & -\delta^+_L/\surd 2
\end{array}\right)
\sim (1,3,1,2), \hspace*{0.2cm}
\bar{\bigtriangleup} =
\left(\begin{array}{cc}
\ \triangle^-_L\surd 2 & \triangle^0_L \\
\ \triangle^{--}_L & -\triangle^-_L/\surd 2
\end{array}\right)
\sim (1,3,1,-2),\nonumber 
\end{equation}
\begin{equation}
\bigtriangleup_c =
\left(\begin{array}{cc}
\ \triangle^-_R/\surd 2 & \triangle^0_R \\
\ \triangle^{--}_R & -\triangle^-_R/\surd 2
\end{array}\right)
\sim (1,1,3,-2), \hspace*{0.2cm}
\bar{\bigtriangleup}_c =
\left(\begin{array}{cc}
\ \delta^+_R/\surd 2 & \delta^{++}_R \\
\ \delta^0_R & -\delta^+_R/\surd 2
\end{array}\right)
\sim (1,1,3,2) \nonumber
\end{equation}
Under the left-right symmetry the fields transform as
\begin{equation}
Q\leftrightarrow Q^*_c,L\leftrightarrow L^*_c,\Phi\leftrightarrow 
\Phi^{\dagger},\bigtriangleup \leftrightarrow \bigtriangleup^*_c, 
\bar{\bigtriangleup}\leftrightarrow \bar{\bigtriangleup^*}_c  \nonumber
\end{equation}
It turns out that left-right symmetry imposes rather strong constraints on
the ground state of this model. It was pointed out by Kuchimanchi and
Mohapatra \cite{Kuchimanchi:1993jg} that there is no spontaneous parity
breaking for this minimal choice of Higgs in the supersymmetric left-right 
model and as such the ground state remains parity symmetric. If parity odd 
singlets are  introduced to break this symmetry \cite{Cvetic:1985zp}, then it was
shown \cite{Kuchimanchi:1993jg} that the charge-breaking vacua have a
lower potential than the charge-preserving vacua and as such the ground
state does not conserve electric charge. Breaking $R$ parity was another
possible solution to this dilemma of breaking parity symmetry. However, if
one wants to prevent proton decay, then one must look for alternative
solutions. One such possible solution is to add two new triplets superfields
$\Omega(1,3,1,0)$, $\Omega_c (1,1,3,0)$ where under parity symmetry
$\Omega \leftrightarrow \Omega_c^*$. This field has been explored
extensively in \cite{Aulakh:1997ba, Aulakh:1997fq, Aulakh:1998nn, Aulakh:1997vc}. 
As worked out in the paper~\cite{Aulakh:1997ba} in this model the $SU(2)_R$ 
breaking takes place in two stages.
\begin{displaymath}
SU(2)_L\otimes SU(2)_R\otimes U(1)_{B-L}\underrightarrow{\langle 
\Omega_c \rangle} SU(2)_L\otimes 
U(1)_R\otimes U(1)_{B-L}\underrightarrow{\langle \bigtriangleup_c \rangle} SU(2)_L\otimes U(1)_Y
\end{displaymath}
The superpotential then takes the form
\begin{eqnarray}
\lefteqn{W=h^{(i)}_l L^T\tau_2 \Phi_i \tau_2 L_c+ h^{(i)}_q Q^T\tau_2 
\Phi_i \tau_2 Q_c+ \iota fL^T\tau_2 \bigtriangleup L +\iota f^*L^T_c 
\tau_2 \bigtriangleup_c L_c} \nonumber \\
&& +m_\triangle \text{Tr}\bigtriangleup \bar{\bigtriangleup}
+m^*_\Delta \text{Tr}\bigtriangleup_c \bar{\bigtriangleup}_c 
+\frac{m_\Omega}{2}\text{Tr}\Omega^2 +\frac{m^*_\Omega}{2} 
\text{Tr}\Omega^2_c \nonumber \\
&& +\mu_{ij}\text{Tr}\tau_2\Phi^T_i\tau_2\Phi_j +a\text{Tr}
\bigtriangleup \Omega \bar{\bigtriangleup}+a^*\text{Tr}
\bigtriangleup_c\Omega_c \bar{\bigtriangleup}_c \nonumber \\
&& +\alpha_{ij}\text{Tr}\Omega \Phi_i\tau_2 \Phi^T_j \tau _2 + 
\alpha^*_{ij}\text{Tr}\Omega_c \Phi^T_i\tau_2 \Phi_j \tau_2
\end{eqnarray}
where $h^{(i)}_{q,l}=h^{(i)\dagger}_{q,l},\mu_{ij}=\mu{ji}=
\mu^*_{ij},\alpha_{ij}=-\alpha_{ji}$ and $f,h$ are symmetric matrices.
It is clear from the 
above superpotential that the theory has no baryon or lepton number violation 
terms. As such  
$R$-parity, defined by $(-1)^{3(B-L)+2S}$, is automatically 
conserved in the SUSYLR model.
In this model the authors arrive at $ M^2_{B-L} = M_{EW} M_R $ which relates different symmetry
breaking scales. However this is achieved under the assumption that $m_{\Omega}$ is of the order of 
electroweak scale. Also the neutrino masses arise in this model through  the so called type II seesaw mechanism
coming from the majorana mass term of the neutrinos. Here we want to explore another possibility of obtaining 
neutrino masses without introducing majorana neutrino masses but by adding a set of singlet fermions which we discuss below.

\section{Minimal Supersymmetric Left-Right Model with Higgs Doublets}
\label{susylr:higgs}
In the present paper we discuss another alternative implementation of supersymmetry
as well as left-right symmetry with a different seesaw mechanism for generating small standard model neutrino masses. Unlike in the model we discussed in the previous section here the 
triplets $\Delta$'s are replaced by four doublets with odd $B-L$ charge.
We show that the ground state of this model breaks parity and also obeys
electromagnetic charge invariance. Also the addition of one parity even singlet $\rho(1,1,1,0)$ per generation will give rise to neutrino masses from so called type III seesaw mechanism. Although type III seesaw was already there in the literature~\cite{Foot:1988aq,Ma:1998dn,Bajc:2006ia} here we incorporate it within a supersymmetric left-right model and talk about the possible phenomenology. Since we want seesaw only in the leptonic sector we assign appropriate quantum numbers to the singlet such that it couples only to leptons and hence may be worth studying from PAMELA~\cite{Adriani:2008zr,Adriani:2008zq} positron excess point of view. The non-supersymmetric version of the left-right model with type-III Seesaw has been worked out in~\cite{FileviezPerez:2008sr}. Here we however consider fermionic singlets instead of triplets(as in conventional type-III seesaw).
The superfields are
\begin{equation}
L(2,1,-1), \quad L_c(1,2,1), \quad \rho (1,1,0), \quad
Q(2,1,\frac{1}{3}),\quad  Q_c(1,2, -\frac{1}{3}) \nonumber
\end {equation}
\begin{equation}
H=
\left(\begin{array}{cc}
\ H^+_{L} \\
\ H^0_{L}/{\surd 2}
\end{array}\right)
\sim (2,1,1), \quad
H_c=
\left(\begin{array}{cc}
\ H^+_{R} \\
\ H^0_{R}/{\surd 2}
\end{array}\right)
\sim (1,2,-1), \nonumber
\end{equation}
\begin{equation}
\bar{H}=
\left(\begin{array}{cc}
\ h^0_{L}/{\surd 2} \\
\ h^-_{L}
\end{array}\right)
\sim (2,1,-1), \quad
\bar{H}_c=
\left(\begin{array}{cc}
\ h^0_{R}/{\surd 2} \\
\ H^-_{R}
\end{array}\right)
\sim (1,2,1), \nonumber
\end{equation}
\begin{equation}
\Phi_1(2,2,0), \quad \Phi_2(2,2,0) \nonumber 
\end{equation}
where the numbers in brackets correspond to the quantum numbers corresponding 
to $ SU(2)_L\times SU(2)_R \times U(1)_{B-L} $. The symmetry breaking pattern 
is 
\begin{eqnarray}
SU(2)_L \times SU(2)_R \times U(1)_{B-L} 
\underrightarrow{\langle H,H_c \rangle} SU(2)_L \times U(1)_{Y} 
\underrightarrow{\langle \Phi \rangle} U(1)_{em}
\end{eqnarray}
The superpotential relevant for the spontaneous parity violation is given as follows
\begin{eqnarray}
\lefteqn{W=h^{(i)}_l L^T\tau_2 \Phi_i \tau_2 L_c+ h^{(i)}_q Q^T\tau_2 
\Phi_i \tau_2 Q_c+ \iota fL^T \tau_2 \rho H +\iota f^*L^T_c 
\tau_2 \rho H_c} \nonumber \\
&& +\mu_{ij}\text{Tr}\tau_2\Phi^T_i\tau_2\Phi_j +M_{\rho}
\rho \rho + f_1(H^T\Phi_i H_c+\bar{H}^T\Phi_i \bar{H}_c)+ \zeta_{ij} \rho\text{Tr}\tau_2\Phi^T_i\tau_2\Phi_j \nonumber \\
&& + m_H H^T \tau_2 \bar{H} +m^*_H H^T_c \tau_2 \bar{H}_c+ \lambda_1 \rho (H^T \tau_2 \bar{H} +  H^T_c \tau_2 \bar{H}_c)
\end{eqnarray}
The corresponding F-conditions are
\begin{eqnarray}
&& -F^*_{\rho} =M_{\rho} \rho + \zeta_{ij} \text{Tr}\tau_2\Phi^T_i\tau_2\Phi_j +\lambda_1 (H^T \tau_2 \bar{H}+H^T_c \tau_2 \bar{H}_c)
 = 0 \nonumber \\
& & -F^*_{H} =f_1\phi_{i}H_{c}+m_{H}\tau_{2}\bar{H}
+ \lambda_1 \rho \tau_{2}\bar{H}=0 \nonumber \\
& & -F^*_{\bar{H}} =f_1\phi_{i}\bar{H}_{c} +m_{H}\tau_{2} H +\lambda_1 \rho \tau_{2}H=0 \nonumber \\
& & -F^*_{H_c} =f_1 \phi_{i}H + m_{H}\tau_{2}\bar{H}_{c}+\lambda_1 \rho \tau_{2}\bar{H}_{c}=0 \nonumber \\
& & -F^*_{\bar{H}_{c}} = f_1\phi_{i}\bar{H}+m_{H}\tau_{2}H_{c}+\lambda_1 \rho \tau_{2}H_{c} =0 \nonumber \\
& & -F^*_{\Phi}=2 \mu \phi + 2 \rho \phi +f_{1}\left( H_{c} H^{T} +\bar{H}_{c}\bar{H}^{T} \right) =0
\end{eqnarray}
Here the neutral fields only acquire any vacuum expectation values(vev) otherwise they 
will lead to breaking of the electromagnetic charge invariance. Also we have neglected the 
slepton and squark fields since they would have zero vev at the scale considered 
in our model. The vacuum of our model should be consistent with the above equations. 
We assign the vev's to the neutral fields:
$\langle \phi^0_{11} \rangle = v_1$, \,$\langle \phi^0_{22} \rangle = v_2$,  \, 
$\langle \phi^0_{12} \rangle = v'_1$,  \,$\langle \phi^0_{21} \rangle = v'_2$,  \,
$ \langle H^0_L \rangle = \langle h^0_L \rangle = v_L$,  \,$\langle H^0_R \rangle = 
\langle h^0_R \rangle = v_R$, \, $\langle \tilde{\rho} \rangle = s $.
The equations of phenomenological relevance are as follows 
\begin{eqnarray}
& & 4 M_{\rho} s + 4(\zeta_{11}v_1 v'_1+\zeta_{12}v'_1 v'_2+\zeta_{12}v_1 v_2+\zeta_{22}v_2 v'_2) +i v_L^2 \lambda_1 +i v_R^2 \lambda_1 =0 \\
& & i v_L \lambda_1 s +f_1 (v'_1+ v_2)v_R +i m_H v_L = 0 \\
& & i v_L \lambda_1 s +f_1 (v_1+ v'_2)v_R +i m_H v_L = 0 \\
& & i v_R \lambda_1 s +f_1 (v'_1+ v_2)v_L +i m^*_H v_R = 0 \\
& & i v_R \lambda_1 s +f_1 (v_1+ v'_2)v_L +i m^*_H v_L = 0 \\
& & v_L v_R f_1 +4 v'_1 (\mu_{11}+\zeta_{11} s) +4 v_2 (\mu_{12}+ \zeta_{12} s) = 0 \\
& & v_L v_R f_1 +4 v_1 (\mu_{11}+\zeta_{11} s) +4 v'_2 (\mu_{12}+ \zeta_{12}s) = 0 \\
& & v_L v_R f_1 +4 v'_1 (\mu_{12}+\zeta_{12} s) +4 v_2 (\mu_{22}+ \zeta_{22} s) = 0 \\
& & v_L v_R f_1 +4 v_1 (\mu_{12}+\zeta_{12}s) +4 v'_2 (\mu_{22}+ \zeta_{22} s) = 0
\end{eqnarray}

From the above relations we can show that parity is broken spontaneously
and at the same time electromagnetic charge is automatically
preserved. Also there is a seesaw between $v_L$ and $v_R$ from the above equations and hence we can arrive at small neutrino masses. However the effective $\mu$ term becomes $ \mu + \zeta s$ which becomes of the order of $v^0_R$, the right handed breaking scale which is generally much higher than the electroweak scale . This will make the standard model Higgs bosons $\phi$ very heavy. To avoid this, as shown in the next section we add two extra triplets with $B-L$ charge zero to the above field content.
\section{SUSYLR model with Higgs doublet as well as triplets}
As we have seen in the last section, keeping the right handed breaking scale higher than the electroweak scale makes the electroweak bosons very heavy. To get rid of this problem we add two extra triplet superfields $\Omega (3,1,0) $ and $\Omega_c(1,3,0)$. These triplets play similar roles in symmetry breaking as in \cite{Aulakh:1997ba, Aulakh:1997fq, Aulakh:1998nn, Aulakh:1997vc} discussed in section~\ref{Aulakh}. Although we include them in our model to get rid if the problem just mentioned, our main focus in on the new phenomenology arising from the singlet superfield. After the inclusion of these extra triplets the superpotential takes the following form
\begin{eqnarray}
\lefteqn{W=h^{(i)}_l L^T\tau_2 \Phi_i \tau_2 L_c+ h^{(i)}_q Q^T\tau_2 
\Phi_i \tau_2 Q_c+ \iota fL^T \tau_2 \rho H +\iota f^*L^T_c 
\tau_2 \rho H_c} \nonumber \\
&& +\mu_{ij}\text{Tr}\tau_2\Phi^T_i\tau_2\Phi_j +\zeta_{ij} \rho\text{Tr}\tau_2\Phi^T_i\tau_2\Phi_j +M_{\rho}
\rho \rho + f_1(H^T\Phi_i H_c+\bar{H}^T\Phi_i \bar{H}_c) \nonumber \\
&& +\alpha_{ij}\text{Tr}\Omega 
\Phi_i\tau_2 \Phi^T_j \tau _2 + \alpha^*_{ij}\text{Tr}\Omega_c \Phi^T_i\tau_2 \Phi_j \tau_2 +
m_H H^T \tau_2 \bar{H} +m^*_H H^T_c \tau_2 \bar{H}_c \nonumber \\
&& +f_3 H^T \tau_2 \Omega \bar{H}+f^*_3 H^T_c\tau_2 \Omega_c \bar{H}_c+ 
\frac{m_{\Omega}}{2}\text{Tr}(\Omega^2 +\Omega^2_c)+\lambda_1 \rho (H^T \tau_2 \bar{H} +  H^T_c \tau_2 \bar{H}_c) \nonumber \\
&& +\iota f_4 L^T \tau_2\Omega H +\iota f_4^*L^T_c \tau_2 \Omega_c H_c + \lambda_2 \rho \text{Tr}( \Omega \Omega + \Omega_c \Omega_c)
\end{eqnarray}
where $h^{(i)}_{q,l}=h^{(i)\dagger}_{q,l},\mu_{ij}=\mu_{ji}=\mu^*_{ij},\zeta_{ij}=\zeta_{ji}=\zeta^*_{ij}$ 
and $f,h$ are symmetric matrices. \\
\indent Now for the spontaneous symmetry breaking purpose we need to 
minimize the scalar potential. In supersymmetric  theories minimization 
of the scalar potential corresponds to $F_{\phi}= 0$, where 
$ F_{\phi}=-\frac{\partial W^\dagger}{\partial \phi^\dagger} $. 
Writing the F-terms and their component forms as shown in appendix \ref{sec:ap1} will determine the vacua of the model which has to be consistent with those equations. Since only neutral fields can acquire vev we assign them their vevs as follows:
$\langle \phi^0_{11} \rangle = v_1$, \,$\langle \phi^0_{22} \rangle = v_2$,  \, 
$\langle \phi^0_{12} \rangle = v'_1$,  \,$\langle \phi^0_{21} \rangle = v'_2$,  \,
$ \langle H^0_L \rangle = \langle h^0_L \rangle = v_L$,  \,$\langle H^0_R \rangle = 
\langle h^0_R \rangle = v_R$,  \,
$\langle \Omega^0_L \rangle = \omega_L \neq 0$, \,$\langle \Omega^0_R \rangle = 
\omega_R$ and $ \langle \tilde{\rho} \rangle = s$.
Here we have assumed $ v'_1, v'_2 \ll v_1, v_2 $ so that the terms containing $v'_1, v'_2$ can be neglected. Under these conditions the component form of the F-terms mentioned in appendix \ref{sec:ap1} become
\begin{eqnarray}
& & 4 \sqrt{2} M_{\rho} s +4 \sqrt{2}(\zeta_{11}v_1 v'_1+\zeta_{12}v'_1 v'_2+\zeta_{12}v_1 v_2+\zeta_{22}v_2 v'_2) \nonumber \\
& &+ i \sqrt{2} \lambda_1 (v^2_L +v^2_R) +\sqrt{2} \lambda_2 \omega^2_R = 0  
\label{F1}  \\
& &
4 \alpha_{12} (v_1 v_2-v'_1 v'_2) + if_3 v^2_L = 0
\label{F2} \\
& &2 \sqrt{2} m_{\Omega} \omega_R + 4 \alpha^*_{12} (v_1 v_2-v'_1 v'_2) + i f_3 v^2_R +4 \sqrt{2} \lambda_2 s \omega_R = 0 
\label{F3} \\
& & \sqrt{2} f_1 (v_2+v'_1) v_R +i \sqrt{2} m_H v_L + i \lambda_1 \sqrt{2} s v_L =0 
\label{F4} \\
& &\sqrt{2} f_1 (v_1+v'_2) v_R +i \sqrt{2} m_H v_L + i \lambda_1 \sqrt{2} s v_L =0
\label{F5}\\
& &\sqrt{2} f_1 (v_2+v'_1) v_L+ i f_3 v_R \omega_R  +i \sqrt{2} m_H v_R + i \lambda_1 \sqrt{2} s v_R =0
\label{F6} \\
& &\sqrt{2} f_1 (v_1+v'_2) v_L+ i f_3 v_R \omega_R  +i \sqrt{2} m_H v_R + i \lambda_1 \sqrt{2} s v_R =0 
\label{F7} \\
& &f_1 v_L v_R +2 \sqrt{2} \alpha'_{12} v_2 \omega_R +4 (\mu_{11}+\zeta_{11}s) v'_1+4 (\mu_{12}+\zeta_{12}s) v_2 = 0
\label{F8} \\
& &f_1 v_L v_R -2 \sqrt{2} \alpha'_{12} v'_2 \omega_R +4 (\mu_{11}+\zeta_{11} s) v_1+4 (\mu_{12}+\zeta_{12} s) v'_2 = 0
\label{F9} \\
& &f_1 v_L v_R -2 \sqrt{2} \alpha'_{12} v'_1 \omega_R +4 (\mu_{12}+\zeta_{12}s) v'_1+4 (\mu_{22}+\zeta_{22}s) v_2 = 0
\label{F10} \\
& &f_1 v_L v_R +2 \sqrt{2} \alpha'_{12} v_1 \omega_R +4 (\mu_{12}+\zeta_{12} s) v_1+4 (\mu_{22}+\zeta_{22} s) v'_2 = 0
\label{F11}
\end{eqnarray}
\subsection{Results}
After getting the above equations relating different vev's we have to scale them in such a way that these relations 
remain consistent. From equation \ref{F2} we have 
\begin{equation}
4 \alpha_{12} v_1 v_2 \simeq  f_3 v^2_L 
\end{equation}
which says that $v_L$ is of the order of GeV since $v_1, v_2$ are of the order of electroweak scale. 
From equations \ref{F4} and \ref{F5}, we will get the seesaw relation like
\begin{equation}
\frac{v_L}{v_R}=\frac{-i\, f_1(v_1+v_2)}{M_H+\lambda_1 \, s} 
\end{equation}
which gives the much needed seesaw between the scales $v_L$ and $v_R$. As we see in the next section this is crucial to generate small neutrino masses. 
We summarize all the energy scales for the consistency of our model in the table \ref{table1}. To keep the effective $\mu$ term at the electroweak scale we set the scale $s$ at the electroweak scale. In that case from equation \ref{F3} it is clear that the scale of $m_{\Omega}$ should be of the order of $v_R$.
\begin{center}
\begin{table}[ht]
\begin{tabular}{|c|c|c|c|}
\hline
Fields          & respective vevs     & Energy scale(in GeV)  & Mass Scales  \\
\hline
$\Omega_c$      &  $\omega_R$         & $10^{12}$             & $m_H$            \\
$H_c$, $\bar{H}_c$    &  $v_R$        & $10^{10}$             & $m_{\Omega}$                    \\ 
$\tilde{\rho}$        &  $s$        &  $10^2$                 &     \\ 
$\Phi_{11}$           &  $v_1$        & $10^{2}$              &                    \\
$H$, $\bar{H}$        &  $v_L$        & 3 $\times 10^{1}$     &                  \\
$\Phi_{22}$           &  $v_2$        & $10^{1}$              &                 \\
\hline
\end{tabular}
\label{table1}
\end{table}
\end{center}
\section{Phenomenology}
\indent The model we have worked out can have various possible phenomenological consequences both from collider as well as astrophysical point of view. Among them , neutrino masses, dark matter candidates and leptogenesis are of primary interests. Neutrino masses arise naturally in this model by so called Type III seesaw mechanism. The neutrino mass matrix in the basis $(\nu, \nu_c, \rho)$ is given by
\begin{equation}
M_{\nu \rho} =
\left(\begin{array}{cccc}
\ 0 & M_D &  F v_L \\
\ M^T_D & 0 & F' v_R\\
\ F v_L & F' v_R  & M_{\rho}
\end{array}\right)
\end{equation}
where $M_D = (\phi^0_{12}h_1+\phi^0_{22}h_2), F = f/\sqrt{2}, F' = f_*/\sqrt{2} $. After orthogonalization we get the following expression for $ \nu $ mass 
\begin{equation}
M_{\nu} = - M_D M_R^{-1} M_D^T - \left( M_D + M_D^T \right)v_{L}/v_{R} 
\label{nu1}
\end{equation}
where 
\begin{equation}
M_R = (F' \,v_{R}) {M_{\rho}}^{-1} (F'^T \,v_{R})
\label{nu2}
\end{equation}
The first term in equation \ref{nu1}  is the usual type I seesaw contribution. The 
second term is the type III seesaw contribution. If the elements of the
matrix $M_{\rho}$ are small compared to those of 
$F' v_{R}$, then it is easy to see from equations \ref{nu1} and \ref{nu2}
that the type I contribution
becomes negligible compared to the type III contribution. This double seesaw was also worked out 
in \cite{FileviezPerez:2008sr} where the right handed neutrino masses come out after integrating out the extra fermionic singlets $\rho$. The smallness of standard model neutrino masses arise naturally by suitable adjustment of the scales $\langle H, H_c \rangle $ and $M_{\rho} $. We can easily get the small $(~eV)$ neutrino masses using the values given in table \ref{table1} in the above expression. \\ 
\indent It should be noted from the neutrino mass matrix that these mass terms allow the mixing of an R-parity odd singlet fermion $\rho$ with an R-parity even neutrino. Note that the superpotential preserves R parity. The mild R parity
violation occuring in the
neutrino mass matrix should be understood as an accidental consequence of
$B-L$ gauge symmetry
breakdown. Apart from these mass terms there are however no vertices in the model which violate R-parity and hence would not lead to dangerous proton decay. Thus the accidental R-parity breaking in our model is phenomenologically acceptable. \\
\indent In supersymmetric models with conserved R-parity the lightest supersymmetric particle is stable and hence a viable dark matter candidate. In the supersymmetric left-right model with gauged B-L which we have discussed in the previous section R-parity is automatically conserved except for the mass terms which mixes neutrino with the singlet fermion. In most SUSY models the neutralinos (the neutral gauginos and higgsinos) are the best dark matter candidates since they satisfy the relic density of dark matter as observed by WMAP (Wilkinson Microwave Anisotropy Probe). Like the minimal SUSYLR model with type 2 seesaw, in our model also we have thirteen neutralinos: linear combination of three neutral gauginos and ten neutral higgsinos. The standard relic abundance analysis can be done for neutralino dark matter using publicly available packages like DarkSUSY\cite{Gondolo:2004sc}, micrOMEGA \cite{Belanger:2006is} etc. We however are not going to do a detailed analysis of neutralino dark matter in our model. Instead of this, we want to focus on some new candidate in our model which can be a dark matter candidate and can overcome the difficulties associated with the usual neutralino as dark matter candidates in existing models. \\
\indent Neutralino dark matter scenario has faced severe problems after experimental results from various indirect detection experiments like PAMELA, ATIC have come out. The PAMELA has observed excess positrons in the cosmic radiation but no excess anti-protons \cite{Adriani:2008zr,Adriani:2008zq}. The positron fraction increases as the energy is varied from 10 GeV to 100 GeV. Although there are many explanations for the observed excess in terms of astrophysical sources \cite{Kobayashi:2003kp,Albright:2003xb}, the connection to dark matter annihilations seems very promising as well as exciting. However the existing dark matter models in particle physics mostly consider neutralino as a dark matter candidate and neutralinos generally annihilate more preferentially into heavier fermions such as quarks due to helicity suppression of majorana fermions. Thus the standard neutralino dark matter candidate won't be able to explain the excess of positrons and simultaneously the lack of antiprotons as observed by the PAMELA. \\
\indent There have been a large number of attempts to explain the positron excess coming from cold dark matter annihilation. Most of the literature talk about hidden gauge force with a light ($\sim$GeV) gauge boson and the dark matter particles annihilate via this new gauge boson exchange to lighter fermions (leptons). Thus annihilations into quark-antiquark pairs are kinematically forbidden. This scenerio has been discussed in detail in \cite{ArkaniHamed:2008qp,ArkaniHamed:2008qn,Katz:2009qq} and the references therein.  However if we can find some particles within existing frameworks which have no direct coupling to quarks and also can eliminate all the channels through which those particles can annihilate into quark-antiquark pairs while keeping the same channels to leptons open we can possibly explain the observed positron excess. But while doing so we have to make sure that the dark matter candidate satisy the observed relic abundance constraints also. That will be slightly non-trivial since in this case the dark matter particle has fewer annihilation channels than the standard neutralino in MSSM which may result in very high relic abundance of dark matter. \\
\begin{figure}[h]
 \centering
 \epsfig{file=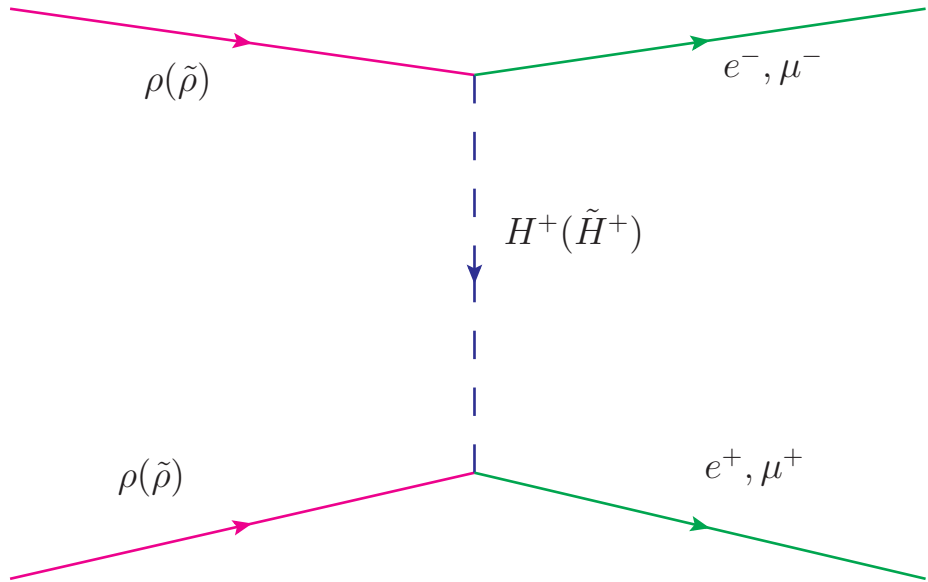}
 \caption{Diagram where the singlets as well as their superpartners annihilate into leptons}
 \label{fig:1}
\end{figure}
\indent In our model we have two such particles which has no annihilation channels to quark-antiquark 
pairs. The singlet fermion $\rho$ and its superpartner $\tilde{\rho}$. The superpartner $\tilde{\rho}$ 
will be naturally protected by R-parity if it is the lightest among the superparticles. However the 
singlet fermion $\rho$ can in principle decay to lighter fermions. But we can incorporate some 
discrete $Z_2$ symmetry under which $\rho \rightarrow -\rho$ and hence it will become a stable 
particle. From the superpotential it is clear that $\rho$ has no direct coupling to quark-antiquarks 
and being a singlet it has no annihilation channels to quark-antiquarks via s-channel exchange of 
gauge bosons. However $\rho$ has direct coupling to leptons. They can annihilate into 
electron-positron pairs through t-channel exchange of higgs boson as shown in \ref{fig:1}. 
The detailed quantitative analysis of the relic abundance of such a dark matter candidate  is a part 
of future work. The calculations regarding the positron excess will go on similar footing as done in 
\cite{Baltz:2002we}. There they considered the case of a right handed neutrino instead of the sterile 
fermion $\rho$ we have here. Although the $\rho $ 's have similar annihilation channels in our model 
like right handed neutrinos have in the model \cite{Baltz:2002we}, the mass mechanism is entirely 
different in the two cases. \\
\indent The leptogenesis mechanism in this model can be very different from the usual SUSYLR model 
due to the absence of Higgs triplets with even B-L where the leptogenesis arises from the doubly 
charged higgs or a heavy neutrino. This has been discussed in detail in  \cite{Albright:2003xb}.
 Apart from all these this new model can have good collider phenomenology in terms of the extra 
fermionic singlets. These singlets can be produced in the collider through processes like 
$ l \bar{l} \rightarrow \rho \rho $ by the channel shown in \ref{fig:1} where the leptons 
$l, \bar{l}$ can come from various intermediate products like photons, Z bosons etc.
\section{Conclusion}
\indent We have implemented type III seesaw in the minimal supersymmetric left right model by 
replacing the B-L even Higgs triplets with B-L odd Higgs doublets and an extra fermionic singlet 
per generation. We have also shown that the minimization of the scalar potential gives us a vacuum 
which does not allow left-right symmetry. Thus spontaneous parity violation can be naturally achieved. 
We have also shown the neutrino mass matrix in our model where the small standard model neutrino 
masses can be obtained by suitable adjustment of various scales of symmetry breaking consistent with 
the F-term conditions. We have also commented on the possible dark matter candidates in the model. 
Since R-parity is conserved in our model except for the mass terms mixing neutrino with the singlet fermion, the neutralinos are of course good dark matter candidates
like in many supersymmetric models. Apart from that the extra fermion $\rho$  as well as its superpartner boson $\tilde{\rho}$ can be more interesting
as dark matter candidates from experiments like PAMELA, ATIC point of  view since they have annihilation
channels only into leptons and not to quarks. We have also mentioned the possible channel which would 
be of interest in the collider aspects of these new singlet fermions.
\section{Acknowledgment}
\indent Sudhanwa Patra would like to thank Prof. Utpal Sarkar for valuable comments and reading the manuscript and also thank the hospitality at Indian Institute of Technology, Gandhinagar, India, where most of the present work was done. Both Debasish Borah and Sudhanwa Patra would like to thank Prof. Urjit A. Yajnik and Prof. S. Umasankar for useful discussions and comments related to the manuscript.
\appendix
\section{F-terms of the SUSYLR model with doublets as well as triplets}\label{sec:ap1}
The superpotential for the model with Higgs doublets as well as triplets is 
\begin{eqnarray}
\lefteqn{W=h^{(i)}_l L^T\tau_2 \Phi_i \tau_2 L_c+ h^{(i)}_q Q^T\tau_2 
\Phi_i \tau_2 Q_c+ \iota fL^T \tau_2 \rho H +\iota f^*L^T_c 
\tau_2 \rho H_c} \nonumber \\
&& +\mu_{ij}\text{Tr}\tau_2\Phi^T_i\tau_2\Phi_j +\zeta_{ij} \rho\text{Tr}\tau_2\Phi^T_i\tau_2\Phi_j +M_{\rho}
\rho \rho + f_1(H^T\Phi_i H_c+\bar{H}^T\Phi_i \bar{H}_c) \nonumber \\
&& +\alpha_{ij}\text{Tr}\Omega 
\Phi_i\tau_2 \Phi^T_j \tau _2 + \alpha^*_{ij}\text{Tr}\Omega_c \Phi^T_i\tau_2 \Phi_j \tau_2 +
m_H H^T \tau_2 \bar{H} +m^*_H H^T_c \tau_2 \bar{H}_c \nonumber \\
&& +f_3 H^T \tau_2 \Omega \bar{H}+f^*_3 H^T_c\tau_2 \Omega_c \bar{H}_c+ 
\frac{m_{\Omega}}{2}\text{Tr}(\Omega^2 +\Omega^2_c)+\lambda_1 \rho (H^T \tau_2 \bar{H} +  H^T_c \tau_2 \bar{H}_c) \nonumber \\
&& +\iota f_4 L^T \tau_2\Omega H +\iota f_4^*L^T_c \tau_2 \Omega_c H_c + \lambda_2 \rho \text{Tr}( \Omega \Omega + \Omega_c \Omega_c) \nonumber
\end{eqnarray}
The corresponding F-terms are 
\begin{eqnarray}
&& -F^*_{\rho} =M_{\rho} \rho +\zeta_{ij} \text{Tr}\tau_2\Phi^T_i\tau_2\Phi_j+ \lambda_1 (H^T \tau_2 \bar{H}+H^T_c \tau_2 \bar{H}_c)
+ \lambda_2 \text{Tr}(\Omega^2 +\Omega^2_c)  = 0 \nonumber \\
& & -F^*_{\Omega}=m_{\Omega}\Omega+ \alpha_{ij} \phi_{i} \tau_{2}\phi^T_{j}\tau_{2} + 
\iota f_3 \tau_{2}H^T \bar{H}+ \lambda_2 \rho \Omega= 0 \nonumber \\
& & -F^*_{\Omega_{c}}=m_{\Omega_c}\Omega_c+\alpha^*_{ij} \phi^T_{i}\tau_{2}\phi_{j}\tau_{2} 
+ \iota f^{*}_3 \tau_{2}H^T_{c} \bar{H}_{c} + \lambda_2 \rho \Omega_c= 0 \nonumber \\
& & -F^*_{H} =f_1\phi_{i}H_{c}+\iota f_3\tau_{2}\Omega \bar{H}+m_{H}\tau_{2}\bar{H}
+ \lambda_1 \rho \tau_{2}\bar{H}=0 \nonumber \\
& & -F^*_{\bar{H}} =f_1\phi_{i}\bar{H}_{c}+f_{3}H^{T}\tau_{2}\Omega
+m_{H}\tau_{2}H +\lambda_1 \rho \tau_{2}H=0 \nonumber \\
& & -F^*_{H_c} =f_1 \phi_{i} H + f^{*}_{3}\tau_{2}\Omega_{c} \bar{H}_{c}
+ m_{H}\tau_{2}\bar{H}_{c}+\lambda_1 \rho \tau_{2}\bar{H}_{c}=0 \nonumber \\
& & -F^*_{\bar{H}_{c}} = f_1\phi_{i}\bar{H}+f^{*}_{3}\tau_{2}\Omega_{c}H_{c}
+m_{H} \tau_{2}H_{c}+\lambda_1 \rho \tau_{2}H_{c} =0 \nonumber \\
& & -F^*_{\Phi}=2 \mu \phi +2 \zeta \rho \phi +\alpha \Omega \tau_{2}\phi^{T} \tau_{2} +
\alpha \Omega \phi +\alpha \Omega_{c} \tau_{2}\phi \tau_{2} +
\alpha \Omega \phi^{T}+f_{1}\left( H_{c}H^{T}+\bar{H}_{c}\bar{H}^{T} \right) =0 \nonumber
\end{eqnarray} 
In terms of the neutral component fields (which only can acquire vacuum expectation values) 
the above equations become
$$ 4 \sqrt{2} M_{\rho} \rho +4 \sqrt{2}(\zeta_{11}v_1 v'_1+\zeta_{12}v'_1 v'_2+\zeta_{12}v_1 v_2+\zeta_{22}v_2 v'_2) $$
$$ + \sqrt{2} \lambda_1 (i H^0_L h^0_L 
+i H^0_R h^0_R) + \sqrt{2} \lambda_2 (\Omega^0_L \Omega^0_L + \Omega^0_R \Omega^0_R)= 0 $$
$$ 2 \sqrt{2} m_{\Omega}\Omega^0_L+4 \phi^0_{11}\phi^0_{22} \alpha_{12}-4 \phi^0_{12}\phi^0_{21} \alpha_{12}
+ i f_3 H^0_L h^0_L+ 4 \sqrt{2} \lambda_2 \rho \Omega^0_L= 0 $$
$$ 2 \sqrt{2} m_{\Omega}\Omega^0_R+4 \phi^0_{11}\phi^0_{22} \alpha^*_{12}-4 \phi^0_{12}\phi^0_{21} \alpha^*_{12}
+ i f_3 H^0_R h^0_R +4 \sqrt{2}  \lambda_2 \rho \Omega^0_R= 0 $$
$$ \sqrt{2} f_1\phi^0_{12}H^0_R +\sqrt{2} f_1 \phi^0_{22}H^0_R  +i f_3 h^0_L \Omega^0_L+ \sqrt{2} i m_H h^0_L
+i \sqrt{2} \lambda_1 \rho h^0_L =0 $$
$$ \sqrt{2} f_1\phi^0_{11} h^0_R +\sqrt{2} f_1\phi^0_{21}h^0_R  +i f_3 H^0_L \Omega^0_L
+\sqrt{2} i m_H H^0_L +i\sqrt{2} \lambda_1 \rho H^0_L =0 $$
$$ \sqrt{2}f_1\phi^0_{12}H^0_L +\sqrt{2}f_1\phi^0_{22}H^0_L  +i f_3 h^0_R \Omega^0_R 
+i \sqrt{2} m_H h^0_R+i \sqrt{2} \lambda_1 \rho h^0_R=0 $$ 
$$ \sqrt{2} f_1\phi^0_{11}h^0_L+\sqrt{2} f_1\phi^0_{21}h^0_L  +i f_3 H^0_R \Omega^0_R +i \sqrt{2} m_H H^0_R +i \sqrt{2}  \lambda_1 H^0_R \rho=0 $$
$$ h^0_L h^0_R f_1 + 2 \sqrt{2} \phi^0_{22} \Omega^0_L \alpha_{12}+2 \sqrt{2} \phi^0_{22}\Omega^0_R \alpha'_{12} + 4 \phi^0_{12} (\mu_{11}+\zeta_{11}s)+4 \phi^0_{22} (\mu_{12}+\zeta_{12} s) = 0 $$
$$ H^0_L H^0_R f_1 -2 \sqrt{2} \phi^0_{21} \Omega^0_L \alpha_{12}-2 \sqrt{2} \phi^0_{21}\Omega^0_R \alpha'_{12} + 4 \phi^0_{11} (\mu_{11}+\zeta_{11} s)+4 \phi^0_{21} (\mu_{12}+\zeta_{12} s) = 0 $$
$$ h^0_L h^0_R f_1 - 2 \sqrt{2} \phi^0_{12} \Omega^0_L \alpha_{12}-2 \sqrt{2} \phi^0_{12}\Omega^0_R \alpha'_{12} + 4 \phi^0_{12} (\mu_{12}+\zeta_{12} s)+4 \phi^0_{22} (\mu_{22}+\zeta_{22} s) = 0 $$
$$ H^0_L H^0_R f_1 +2 \sqrt{2} \phi^0_{11} \Omega^0_L \alpha_{12}+2 \sqrt{2} \phi^0_{11}\Omega^0_R \alpha'_{12} + 4 \phi^0_{11} (\mu_{12}+\zeta_{12} s)+4 \phi^0_{21} (\mu_{22}+\zeta_{22} s) = 0 $$
where we have neglected the squarks as well slepton vevs. The vacua can be obtained by suitable adjustment of the vacuum expectation values of the fields in the above equations.

\begin{thebibliography}{34}
\expandafter\ifx\csname natexlab\endcsname\relax\def\natexlab#1{#1}\fi
\expandafter\ifx\csname bibnamefont\endcsname\relax
  \def\bibnamefont#1{#1}\fi
\expandafter\ifx\csname bibfnamefont\endcsname\relax
  \def\bibfnamefont#1{#1}\fi
\expandafter\ifx\csname citenamefont\endcsname\relax
  \def\citenamefont#1{#1}\fi
\expandafter\ifx\csname url\endcsname\relax
  \def\url#1{\texttt{#1}}\fi
\expandafter\ifx\csname urlprefix\endcsname\relax\def\urlprefix{URL }\fi
\providecommand{\bibinfo}[2]{#2}
\providecommand{\eprint}[2][]{\url{#2}}

\bibitem[{\citenamefont{Pati and Salam}(1974)}]{Pati:1974yy}
\bibinfo{author}{\bibfnamefont{J.~C.} \bibnamefont{Pati}} \bibnamefont{and}
  \bibinfo{author}{\bibfnamefont{A.}~\bibnamefont{Salam}},
  \bibinfo{journal}{Phys. Rev.} \textbf{\bibinfo{volume}{D10}},
  \bibinfo{pages}{275} (\bibinfo{year}{1974}).

\bibitem[{\citenamefont{Mohapatra and Pati}(1975)}]{Mohapatra:1974gc}
\bibinfo{author}{\bibfnamefont{R.~N.} \bibnamefont{Mohapatra}}
  \bibnamefont{and} \bibinfo{author}{\bibfnamefont{J.~C.} \bibnamefont{Pati}},
  \bibinfo{journal}{Phys. Rev.} \textbf{\bibinfo{volume}{D11}},
  \bibinfo{pages}{2558} (\bibinfo{year}{1975}).

\bibitem[{\citenamefont{Senjanovic and Mohapatra}(1975)}]{Senjanovic:1975rk}
\bibinfo{author}{\bibfnamefont{G.}~\bibnamefont{Senjanovic}} \bibnamefont{and}
  \bibinfo{author}{\bibfnamefont{R.~N.} \bibnamefont{Mohapatra}},
  \bibinfo{journal}{Phys. Rev.} \textbf{\bibinfo{volume}{D12}},
  \bibinfo{pages}{1502} (\bibinfo{year}{1975}).

\bibitem[{\citenamefont{Mohapatra and Marshak}(1980)}]{Mohapatra:1980qe}
\bibinfo{author}{\bibfnamefont{R.~N.} \bibnamefont{Mohapatra}}
  \bibnamefont{and} \bibinfo{author}{\bibfnamefont{R.~E.}
  \bibnamefont{Marshak}}, \bibinfo{journal}{Phys. Rev. Lett.}
  \textbf{\bibinfo{volume}{44}}, \bibinfo{pages}{1316} (\bibinfo{year}{1980}).

\bibitem[{\citenamefont{Deshpande et~al.}(1991)\citenamefont{Deshpande, Gunion,
  Kayser, and Olness}}]{Deshpande:1990ip}
\bibinfo{author}{\bibfnamefont{N.~G.} \bibnamefont{Deshpande}},
  \bibinfo{author}{\bibfnamefont{J.~F.} \bibnamefont{Gunion}},
  \bibinfo{author}{\bibfnamefont{B.}~\bibnamefont{Kayser}}, \bibnamefont{and}
  \bibinfo{author}{\bibfnamefont{F.~I.} \bibnamefont{Olness}},
  \bibinfo{journal}{Phys. Rev.} \textbf{\bibinfo{volume}{D44}},
  \bibinfo{pages}{837} (\bibinfo{year}{1991}).

\bibitem[{\citenamefont{Fukuda et~al.}(2001)}]{Fukuda:2001nk}
\bibinfo{author}{\bibfnamefont{S.}~\bibnamefont{Fukuda}} \bibnamefont{et~al.}
  (\bibinfo{collaboration}{Super-Kamiokande}), \bibinfo{journal}{Phys. Rev.
  Lett.} \textbf{\bibinfo{volume}{86}}, \bibinfo{pages}{5656}
  (\bibinfo{year}{2001}), \eprint{hep-ex/0103033}.

\bibitem[{\citenamefont{Ahmad et~al.}(2002{\natexlab{a}})}]{Ahmad:2002jz}
\bibinfo{author}{\bibfnamefont{Q.~R.} \bibnamefont{Ahmad}} \bibnamefont{et~al.}
  (\bibinfo{collaboration}{SNO}), \bibinfo{journal}{Phys. Rev. Lett.}
  \textbf{\bibinfo{volume}{89}}, \bibinfo{pages}{011301}
  (\bibinfo{year}{2002}{\natexlab{a}}), \eprint{nucl-ex/0204008}.

\bibitem[{\citenamefont{Ahmad et~al.}(2002{\natexlab{b}})}]{Ahmad:2002ka}
\bibinfo{author}{\bibfnamefont{Q.~R.} \bibnamefont{Ahmad}} \bibnamefont{et~al.}
  (\bibinfo{collaboration}{SNO}), \bibinfo{journal}{Phys. Rev. Lett.}
  \textbf{\bibinfo{volume}{89}}, \bibinfo{pages}{011302}
  (\bibinfo{year}{2002}{\natexlab{b}}), \eprint{nucl-ex/0204009}.

\bibitem[{\citenamefont{Bahcall and Pena-Garay}(2004)}]{Bahcall:2004mz}
\bibinfo{author}{\bibfnamefont{J.~N.} \bibnamefont{Bahcall}} \bibnamefont{and}
  \bibinfo{author}{\bibfnamefont{C.}~\bibnamefont{Pena-Garay}},
  \bibinfo{journal}{New J. Phys.} \textbf{\bibinfo{volume}{6}},
  \bibinfo{pages}{63} (\bibinfo{year}{2004}), \eprint{hep-ph/0404061}.

\bibitem[{\citenamefont{Minkowski}(1977)}]{Minkowski:1977sc}
\bibinfo{author}{\bibfnamefont{P.}~\bibnamefont{Minkowski}},
  \bibinfo{journal}{Phys. Lett.} \textbf{\bibinfo{volume}{B67}},
  \bibinfo{pages}{421} (\bibinfo{year}{1977}).

\bibitem[{\citenamefont{Gell-Mann et~al.}(1980)\citenamefont{Gell-Mann, Ramond,
  and Slansky}}]{GellMann:1980vs}
\bibinfo{author}{\bibfnamefont{M.}~\bibnamefont{Gell-Mann}},
  \bibinfo{author}{\bibfnamefont{P.}~\bibnamefont{Ramond}}, \bibnamefont{and}
  \bibinfo{author}{\bibfnamefont{R.}~\bibnamefont{Slansky}}
  (\bibinfo{year}{1980}), \bibinfo{note}{print-80-0576 (CERN)}.

\bibitem[{\citenamefont{Yanagida}(1979)}]{Yanagida:1979as}
\bibinfo{author}{\bibfnamefont{T.}~\bibnamefont{Yanagida}}
  (\bibinfo{year}{1979}), \bibinfo{note}{in Proceedings of the Workshop on the
  Baryon Number of the Universe and Unified Theories, Tsukuba, Japan, 13-14 Feb
  1979}.

\bibitem[{\citenamefont{Mohapatra and Senjanovic}(1980)}]{Mohapatra:1979ia}
\bibinfo{author}{\bibfnamefont{R.~N.} \bibnamefont{Mohapatra}}
  \bibnamefont{and}
  \bibinfo{author}{\bibfnamefont{G.}~\bibnamefont{Senjanovic}},
  \bibinfo{journal}{Phys. Rev. Lett.} \textbf{\bibinfo{volume}{44}},
  \bibinfo{pages}{912} (\bibinfo{year}{1980}).

\bibitem[{\citenamefont{Patra et~al.}(2009)\citenamefont{Patra, Sarkar, Sarkar,
  and Yajnik}}]{Patra:2009wc}
\bibinfo{author}{\bibfnamefont{S.}~\bibnamefont{Patra}},
  \bibinfo{author}{\bibfnamefont{A.}~\bibnamefont{Sarkar}},
  \bibinfo{author}{\bibfnamefont{U.}~\bibnamefont{Sarkar}}, \bibnamefont{and}
  \bibinfo{author}{\bibfnamefont{U.}~\bibnamefont{Yajnik}},
  \bibinfo{journal}{Phys. Lett.} \textbf{\bibinfo{volume}{B679}},
  \bibinfo{pages}{386} (\bibinfo{year}{2009}), \eprint{0905.3220}.

\bibitem[{\citenamefont{Aulakh et~al.}(1998{\natexlab{a}})\citenamefont{Aulakh,
  Melfo, and Senjanovic}}]{Aulakh:1998nn}
\bibinfo{author}{\bibfnamefont{C.~S.} \bibnamefont{Aulakh}},
  \bibinfo{author}{\bibfnamefont{A.}~\bibnamefont{Melfo}}, \bibnamefont{and}
  \bibinfo{author}{\bibfnamefont{G.}~\bibnamefont{Senjanovic}},
  \bibinfo{journal}{Phys. Rev.} \textbf{\bibinfo{volume}{D57}},
  \bibinfo{pages}{4174} (\bibinfo{year}{1998}{\natexlab{a}}),
  \eprint{hep-ph/9707256}.

\bibitem[{\citenamefont{Aulakh et~al.}(1997)\citenamefont{Aulakh, Benakli, and
  Senjanovic}}]{Aulakh:1997ba}
\bibinfo{author}{\bibfnamefont{C.~S.} \bibnamefont{Aulakh}},
  \bibinfo{author}{\bibfnamefont{K.}~\bibnamefont{Benakli}}, \bibnamefont{and}
  \bibinfo{author}{\bibfnamefont{G.}~\bibnamefont{Senjanovic}},
  \bibinfo{journal}{Phys. Rev. Lett.} \textbf{\bibinfo{volume}{79}},
  \bibinfo{pages}{2188} (\bibinfo{year}{1997}), \eprint{hep-ph/9703434}.

\bibitem[{\citenamefont{Kuchimanchi and Mohapatra}(1993)}]{Kuchimanchi:1993jg}
\bibinfo{author}{\bibfnamefont{R.}~\bibnamefont{Kuchimanchi}} \bibnamefont{and}
  \bibinfo{author}{\bibfnamefont{R.~N.} \bibnamefont{Mohapatra}},
  \bibinfo{journal}{Phys. Rev.} \textbf{\bibinfo{volume}{D48}},
  \bibinfo{pages}{4352} (\bibinfo{year}{1993}), \eprint{hep-ph/9306290}.

\bibitem[{\citenamefont{Cvetic}(1985)}]{Cvetic:1985zp}
\bibinfo{author}{\bibfnamefont{M.}~\bibnamefont{Cvetic}},
  \bibinfo{journal}{Phys. Lett.} \textbf{\bibinfo{volume}{B164}},
  \bibinfo{pages}{55} (\bibinfo{year}{1985}).

\bibitem[{\citenamefont{Aulakh et~al.}(1998{\natexlab{b}})\citenamefont{Aulakh,
  Melfo, Rasin, and Senjanovic}}]{Aulakh:1997fq}
\bibinfo{author}{\bibfnamefont{C.~S.} \bibnamefont{Aulakh}},
  \bibinfo{author}{\bibfnamefont{A.}~\bibnamefont{Melfo}},
  \bibinfo{author}{\bibfnamefont{A.}~\bibnamefont{Rasin}}, \bibnamefont{and}
  \bibinfo{author}{\bibfnamefont{G.}~\bibnamefont{Senjanovic}},
  \bibinfo{journal}{Phys. Rev.} \textbf{\bibinfo{volume}{D58}},
  \bibinfo{pages}{115007} (\bibinfo{year}{1998}{\natexlab{b}}),
  \eprint{hep-ph/9712551}.

\bibitem[{\citenamefont{Aulakh}(1997)}]{Aulakh:1997vc}
\bibinfo{author}{\bibfnamefont{C.~S.} \bibnamefont{Aulakh}}
  (\bibinfo{year}{1997}), \eprint{hep-ph/9803461}.

\bibitem[{\citenamefont{Foot et~al.}(1989)\citenamefont{Foot, Lew, He, and
  Joshi}}]{Foot:1988aq}
\bibinfo{author}{\bibfnamefont{R.}~\bibnamefont{Foot}},
  \bibinfo{author}{\bibfnamefont{H.}~\bibnamefont{Lew}},
  \bibinfo{author}{\bibfnamefont{X.~G.} \bibnamefont{He}}, \bibnamefont{and}
  \bibinfo{author}{\bibfnamefont{G.~C.} \bibnamefont{Joshi}},
  \bibinfo{journal}{Z. Phys.} \textbf{\bibinfo{volume}{C44}},
  \bibinfo{pages}{441} (\bibinfo{year}{1989}).

\bibitem[{\citenamefont{Ma}(1998)}]{Ma:1998dn}
\bibinfo{author}{\bibfnamefont{E.}~\bibnamefont{Ma}}, \bibinfo{journal}{Phys.
  Rev. Lett.} \textbf{\bibinfo{volume}{81}}, \bibinfo{pages}{1171}
  (\bibinfo{year}{1998}), \eprint{hep-ph/9805219}.

\bibitem[{\citenamefont{Bajc and Senjanovic}(2007)}]{Bajc:2006ia}
\bibinfo{author}{\bibfnamefont{B.}~\bibnamefont{Bajc}} \bibnamefont{and}
  \bibinfo{author}{\bibfnamefont{G.}~\bibnamefont{Senjanovic}},
  \bibinfo{journal}{JHEP} \textbf{\bibinfo{volume}{08}}, \bibinfo{pages}{014}
  (\bibinfo{year}{2007}), \eprint{hep-ph/0612029}.

\bibitem[{\citenamefont{Adriani et~al.}(2009{\natexlab{a}})}]{Adriani:2008zr}
\bibinfo{author}{\bibfnamefont{O.}~\bibnamefont{Adriani}} \bibnamefont{et~al.}
  (\bibinfo{collaboration}{PAMELA}), \bibinfo{journal}{Nature}
  \textbf{\bibinfo{volume}{458}}, \bibinfo{pages}{607}
  (\bibinfo{year}{2009}{\natexlab{a}}), \eprint{0810.4995}.

\bibitem[{\citenamefont{Adriani et~al.}(2009{\natexlab{b}})}]{Adriani:2008zq}
\bibinfo{author}{\bibfnamefont{O.}~\bibnamefont{Adriani}} \bibnamefont{et~al.},
  \bibinfo{journal}{Phys. Rev. Lett.} \textbf{\bibinfo{volume}{102}},
  \bibinfo{pages}{051101} (\bibinfo{year}{2009}{\natexlab{b}}),
  \eprint{0810.4994}.

\bibitem[{\citenamefont{Fileviez~Perez}(2009)}]{FileviezPerez:2008sr}
\bibinfo{author}{\bibfnamefont{P.}~\bibnamefont{Fileviez~Perez}},
  \bibinfo{journal}{JHEP} \textbf{\bibinfo{volume}{03}}, \bibinfo{pages}{142}
  (\bibinfo{year}{2009}), \eprint{0809.1202}.

\bibitem[{\citenamefont{Gondolo et~al.}(2004)}]{Gondolo:2004sc}
\bibinfo{author}{\bibfnamefont{P.}~\bibnamefont{Gondolo}} \bibnamefont{et~al.},
  \bibinfo{journal}{JCAP} \textbf{\bibinfo{volume}{0407}}, \bibinfo{pages}{008}
  (\bibinfo{year}{2004}), \eprint{astro-ph/0406204}.

\bibitem[{\citenamefont{Belanger et~al.}(2007)\citenamefont{Belanger, Boudjema,
  Pukhov, and Semenov}}]{Belanger:2006is}
\bibinfo{author}{\bibfnamefont{G.}~\bibnamefont{Belanger}},
  \bibinfo{author}{\bibfnamefont{F.}~\bibnamefont{Boudjema}},
  \bibinfo{author}{\bibfnamefont{A.}~\bibnamefont{Pukhov}}, \bibnamefont{and}
  \bibinfo{author}{\bibfnamefont{A.}~\bibnamefont{Semenov}},
  \bibinfo{journal}{Comput. Phys. Commun.} \textbf{\bibinfo{volume}{176}},
  \bibinfo{pages}{367} (\bibinfo{year}{2007}), \eprint{hep-ph/0607059}.

\bibitem[{\citenamefont{Kobayashi et~al.}(2004)\citenamefont{Kobayashi, Komori,
  Yoshida, and Nishimura}}]{Kobayashi:2003kp}
\bibinfo{author}{\bibfnamefont{T.}~\bibnamefont{Kobayashi}},
  \bibinfo{author}{\bibfnamefont{Y.}~\bibnamefont{Komori}},
  \bibinfo{author}{\bibfnamefont{K.}~\bibnamefont{Yoshida}}, \bibnamefont{and}
  \bibinfo{author}{\bibfnamefont{J.}~\bibnamefont{Nishimura}},
  \bibinfo{journal}{Astrophys. J.} \textbf{\bibinfo{volume}{601}},
  \bibinfo{pages}{340} (\bibinfo{year}{2004}), \eprint{astro-ph/0308470}.

\bibitem[{\citenamefont{Albright and Barr}(2004)}]{Albright:2003xb}
\bibinfo{author}{\bibfnamefont{C.~H.} \bibnamefont{Albright}} \bibnamefont{and}
  \bibinfo{author}{\bibfnamefont{S.~M.} \bibnamefont{Barr}},
  \bibinfo{journal}{Phys. Rev.} \textbf{\bibinfo{volume}{D69}},
  \bibinfo{pages}{073010} (\bibinfo{year}{2004}), \eprint{hep-ph/0312224}.

\bibitem[{\citenamefont{Arkani-Hamed and Weiner}(2008)}]{ArkaniHamed:2008qp}
\bibinfo{author}{\bibfnamefont{N.}~\bibnamefont{Arkani-Hamed}}
  \bibnamefont{and} \bibinfo{author}{\bibfnamefont{N.}~\bibnamefont{Weiner}},
  \bibinfo{journal}{JHEP} \textbf{\bibinfo{volume}{12}}, \bibinfo{pages}{104}
  (\bibinfo{year}{2008}), \eprint{0810.0714}.

\bibitem[{\citenamefont{Arkani-Hamed et~al.}(2009)\citenamefont{Arkani-Hamed,
  Finkbeiner, Slatyer, and Weiner}}]{ArkaniHamed:2008qn}
\bibinfo{author}{\bibfnamefont{N.}~\bibnamefont{Arkani-Hamed}},
  \bibinfo{author}{\bibfnamefont{D.~P.} \bibnamefont{Finkbeiner}},
  \bibinfo{author}{\bibfnamefont{T.~R.} \bibnamefont{Slatyer}},
  \bibnamefont{and} \bibinfo{author}{\bibfnamefont{N.}~\bibnamefont{Weiner}},
  \bibinfo{journal}{Phys. Rev.} \textbf{\bibinfo{volume}{D79}},
  \bibinfo{pages}{015014} (\bibinfo{year}{2009}), \eprint{0810.0713}.

\bibitem[{\citenamefont{Katz and Sundrum}(2009)}]{Katz:2009qq}
\bibinfo{author}{\bibfnamefont{A.}~\bibnamefont{Katz}} \bibnamefont{and}
  \bibinfo{author}{\bibfnamefont{R.}~\bibnamefont{Sundrum}}
  (\bibinfo{year}{2009}), \eprint{0902.3271}.

\bibitem[{\citenamefont{Baltz and Bergstrom}(2003)}]{Baltz:2002we}
\bibinfo{author}{\bibfnamefont{E.~A.} \bibnamefont{Baltz}} \bibnamefont{and}
  \bibinfo{author}{\bibfnamefont{L.}~\bibnamefont{Bergstrom}},
  \bibinfo{journal}{Phys. Rev.} \textbf{\bibinfo{volume}{D67}},
  \bibinfo{pages}{043516} (\bibinfo{year}{2003}), \eprint{hep-ph/0211325}.

\end{thebibliography}

\end{document}